\documentclass[prl,a4paper,twocolumn,superscriptaddress,groupedaddress]{revtex4}  
\usepackage{graphicx}  
\usepackage{dcolumn}   
\usepackage{bm}        
\usepackage{amssymb}   

\begin{document}

\date{\today}

\title{
%
%
%

Corners in soft solids behave as defects in crystals

}



\author{Robin Masurel}
\address{Laboratoire Mati\`{e}re et Syst\`{e}mes Complexes, Universit\'{e} Paris Diderot, CNRS UMR 7057, Sorbonne Paris Cit\'{e},
10 Rue A. Domon et L. Duquet, F-75013 Paris, France}
\author{Matthieu Roch\'{e}}
\address{Laboratoire Mati\`{e}re et Syst\`{e}mes Complexes, Universit\'{e} Paris Diderot, CNRS UMR 7057, Sorbonne Paris Cit\'{e},
10 Rue A. Domon et L. Duquet, F-75013 Paris, France}
\author{Laurent Limat}
\address{Laboratoire Mati\`{e}re et Syst\`{e}mes Complexes, Universit\'{e} Paris Diderot, CNRS UMR 7057, Sorbonne Paris Cit\'{e},
10 Rue A. Domon et L. Duquet, F-75013 Paris, France}
\author{Ioan Ionescu}
\address{Laboratoire des Sciences des Proc\'{e}d\'{e}s et des Mat\'{e}riaux, Universit\'{e} Paris 13, CNRS UPR 3407, Sorbonne Paris Cit\'{e}, 99 Avenue J.-B. Clement, F-93430 Villetaneuse, France}
\author{Julien Dervaux}
\email{julien.dervaux@univ-paris-diderot.fr}
\address{Laboratoire Mati\`{e}re et Syst\`{e}mes Complexes, Universit\'{e} Paris Diderot, CNRS UMR 7057, Sorbonne Paris Cit\'{e},
10 Rue A. Domon et L. Duquet, F-75013 Paris, France}


%

\begin{abstract}
All phases of matter, solid, liquid or gas, present some excess energy, compared to their bulk, at their interfaces with other materials. This excess of energy, known as the surface energy, is a fundamental property of matter and is involved in virtually all interface problems in science, from the shape of bubbles, crystals and biological cells to the delicate motion of some insects on water or the fluttering of red blood cells. Because of their high cohesive internal energies, the surface energies of solids differ fundamentally from those of fluids and depend on the solid deformations. This effect, known as the Shuttleworth effect, is well established for metals but is highly debated for amorphous materials such as glasses, elastomers or biological tissues with recent experimental results yielding strictly opposite conclusions with regards to its very existence. Using a combination of analytical results and numerical simulations, we show in this paper that those seemingly opposite results can be reconciled due to the existence of an analog of the Peach-Koehler force acting on the elastocapillary ridge and conclude that: i) there is no large Shuttleworth effect in soft elastomers and ii) the Neumann construction does not hold in elastowetting.
\end{abstract}

\maketitle

As noted by Gibbs \cite{gibbs06}, the surface energy of an interface $\gamma$, defined as the energy required to create a unit of area by a \textit{cleaving} process, differs conceptually from the surface tension $\Upsilon$ of the same interface, which is defined as the force required to create a unit of surface through the \textit{stretching} of this interface. This conceptual difference is of no consequences for fluids as molecules rearrange themselves upon stretching so as to maintain a constant intermolecular distance, such that $\gamma =\Upsilon$ for fluids \cite{cammarata94}. In sharp contrast, molecules in a perfectly elastic solid cannot rearrange themselves and the stretching of a material therefore alters the intermolecular distance such that, in all generality, $\gamma \neq \Upsilon$. However, those two quantities are not independent and are related through the Shuttleworth equation \cite{shuttleworth50}: $\Upsilon(\lambda) = \gamma(\lambda) + \partial \gamma/\partial \lambda$ where $\lambda$ is the stretch parallel to the interface. A model-free determination of the magnitude of the Shuttleworth effect, ie the difference $\Upsilon(\lambda) - \gamma(\lambda)$, is a difficult task in general as there is no direct way to measure the surface stresses of solids. Indirect measurements, based on phonons dispersion or surface-stress induced changes in atomic structure of nanoscale particles, critically depend on models calculations to extract absolute values of surface stresses. Thus, because the atomic structure of metals is very well characterized, reliable measurements for $\Upsilon(\lambda)$ and  $\gamma(\lambda)$ have been obtained for various metals and alloys \cite{cammarata94,payne89,fiorentini93,ibach97,muller04}. 

\begin{figure}[t]
\begin{center}
\includegraphics[width=\columnwidth]{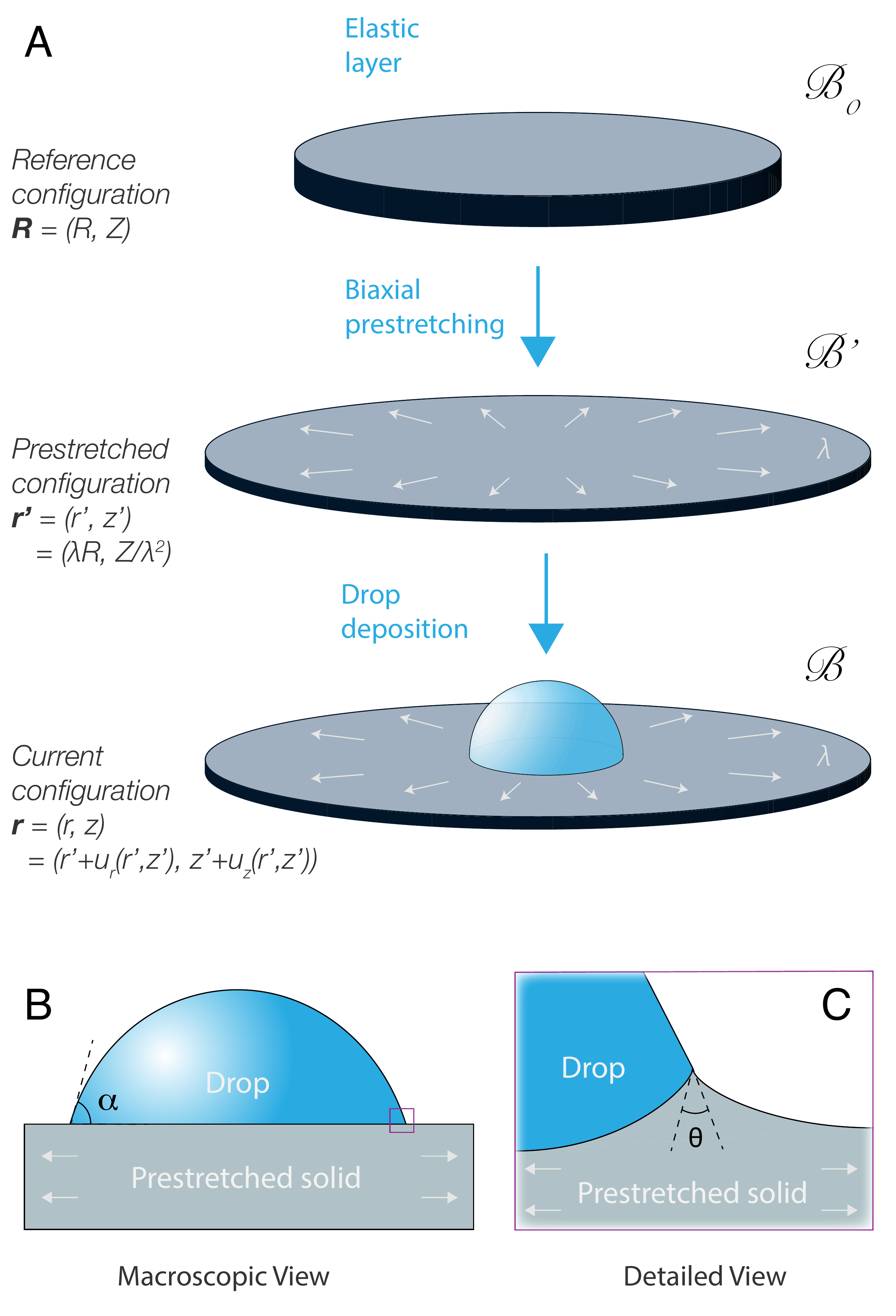}
\caption{{\footnotesize Schematic representation and notations for the problem under consideration. A: a flat layer with initial thickness $H$ and infinite lateral dimensions is first biaxially stretched. A drop is then deposited on this stretched surface and further deforms the elastic layer. B: The shape of the drop depends on the macroscopic contact angle $\alpha$. C: below the contact line, a ridge is formed with opening angle $\theta$.}}
\label{fig:fig1}
\end{center}
\end{figure}

For amorphous materials however, there is a striking lack of experimental data and no consensus has been reached even with regards to the very existence of the Shuttleworth effect for such materials \cite{xu17,schulman18}. In the light of this gap, it has been suggested very recently that the physics of wetting, i.e the interplay between liquid drops and solid surfaces, might shed light on this matter because both the deformation of the solid by surface tension as well as the equilibrium configuration of the drop critically depend on the surface energy of the solid \cite{weijs14,andreotti16a,andreotti16b}. In this line of thought, Schulman et al \cite{schulman18} have measured the macroscopic contact angles $\alpha$ of various liquids on strained glassy and elastomeric materials (see Fig.\ref{fig:fig1} A and B) and concluded that glasses do exhibit strain-dependent surface energies while polymeric materials do not. Building upon the rapidly developing field of elastowetting \cite{limat12,style12,style13,bostwick14,dervaux15,hui14,hui15,mora15,style16}, Xu et al have focused, in another experiment \cite{xu17}, on the ridge formed at the free surface of a soft elastomer below the contact line of a liquid drop, as illustrated in Fig1-C. According to the linear theory of elastowetting the opening angle $\theta$ of this ridge is given by $\theta = \pi - \gamma_{\ell}/\gamma_s$ and is thus solely a function of the ratio between the surface energy of the drop $\gamma_{\ell}$ and that of the solid $\gamma_s$. This elegant result is typically interpreted as the linear approximation of the Neumann construction $\theta = \pi - 2\mbox{arcsin}(\gamma_{\ell}/2\gamma_s)$ that rules the equilibrium of liquid drops on liquid layers. When the elastomeric layer was subject to a biaxial stretch of magnitude $\lambda$ however, they observed an opening of this ridge, i.e and increase in $\theta$ with $\lambda$. This result lead the authors to conclude to the existence of a Shuttleworth effect ($\partial \gamma_s/\partial \lambda \neq 0$) in elastomers. Furthermore, this effect was not small. On the contrary, it was found that the surface energy of elastomers doubled for a mere stretching of $17\%$ of the elastomeric layer, an effect 5 to 10 times larger than that of metals and in complete opposition with the experimental data of Schulman et al \cite{schulman18}.

As we shall see shortly, those seemingly opposite results can in fact be reconciled. First, it should be noted that the critical assumption allowing to conclude on the existence of the Shuttleworth effect in elastomers is the validity of the Neumann construction at the triple line. This result, which stems from the linear elastowetting theory, relies on the hypothesis that the deformations of the solid are small enough such that higher-order terms in the elastic and surface energy densities can be neglected. Obviously, this hypothesis is never verified when the elastic layer is pre-stretched with a finite strain before the deposition of the drop. Furthermore, even in the absence of any pre-stretching, a necessary condition for this hypothesis to hold is that the experimental system under consideration verifies $\gamma_{\ell}/2\gamma_s \ll 1$. This is however never true in currently available experiments where the value of the ratio $\gamma_{\ell}/2\gamma_s$ is typically in the range $0.5-0.9$. The goal of this paper is thus to investigate the elastowetting problem in the nonlinear range in order to answer the following two questions: (i) \textit{does the Neumann construction hold at experimentally relevant finite value of $\gamma_{\ell}/2\gamma_s$ ?} and (ii) \textit{does the Neumann construction hold when the elastic layer is initially prestretched before the deposition of the drop, at experimentally relevant finite values of  $\gamma_{\ell}/2\gamma_s$ ?}

In order to answer those two questions, we will consider an incompressible nonlinearly elastic (Neo-Hookean) layer with a liquid-like surface tension $\gamma_s$ (i.e without any Shuttleworth effect) that is prestretched with a finite strain $\lambda$ before the deposition of a drop with arbitrary surface energy $\gamma_{\ell}$ at its free surface. In the following we will first derive the energy functional of the system and explore the consequence of the stationarity condition. In particular, the nonlinear configurational force balance holding at the triple line will be discussed, showing that a Peach-Koehler force \cite{peach50} acts at the tip of the elastocapilary ridge. Then the standard linear elastowetting theory will be extended to account for the existence of the finite prestretch, i.e the equilibrium equations will be linearized around this finite deformation and the so-called incremental solution will be obtained \cite{ogden84}, valid for arbitrary $\lambda$ but small $\gamma_{\ell}/2\gamma_s$. Next, the full energy functional will be minimized numerically in order to analyze the general case of arbitrary $\lambda$ and $\gamma_{\ell}/2\gamma_s$. We will then compare the incremental (analytical) and fully nonlinear (numerical) solution and answers to the two questions above will be provided. Finally, some comparisons will be made with available experimental data and further approximate analytical results appropriate for the nonlinear regime will be derived.

\bigskip

\noindent {\bf Problem formulation.} We now turn to the formulation of the problem and consider an incompressible elastic body which, in the reference configuration $\mathcal{B}_0$, is a flat layer with initial thickness $H$ and infinite lateral dimensions such that the reference configuration is described in cylindrical coordinates $(R,Z)$ by the region $0 \leq R < \infty$ and $0 \leq Z \leq H$.  This elastic layer is first biaxially stretched such that a material point with position $\mathbf{R} = (R,Z)$ is mapped to a position $\mathbf{r'} = (r', z') = (\lambda R,Z/\lambda^2)$ in the prestretched configuration $\mathcal{B'}$ as a consequence of the incompressibility constraint. The thickness of the prestretched layer is thus $h=H/\lambda^2$. A local description of this deformation is provided by the deformation tensor $\bf{F'} = \partial \bf{r'} / \partial \bf{R}$  In a second step, a drop is deposited at the free surface of the prestretched layer and induces an additional deformation, superposed on the previous finite deformation. There exists therefore another deformation field that maps a point with coordinates $\mathbf{r'}$ in the prestretched configuration $\mathcal{B'}$ to a position $\mathbf{r} = \mathbf{r'} + \mathbf{u(\mathbf{r'})} = (r' + u_r(r',z'), z' + u_z(r',z')) $ in the current configuration $\mathcal{B}$.  Note that the deformation from $\mathcal{B'}$ to $\mathcal{B}$ is expressed in the prestretched coordinates $(r',z')$. The deformation tensor $\bf{F} = \partial \bf{r} / \partial \bf{R}$ is a local description of the overall deformation process. We focus here on homogeneous isotropic incompressible Neo-Hookean, with a strain energy density $\mathcal{W}_e = \frac{\mu}{2} (\mbox{Tr}\mathbf{F}^T\mathbf{F}-3)$. In addition to the elastic energy, we will assume that the system also has a constant, liquid-like, surface energy density $\mathcal{W}_s = \gamma_s$ associated with the free surface in the current configuration $\mathcal{B}$. Furthermore, the incompressibility constraint $ \mbox{det}\, \mathbf{F} = 1$ can be accounted for by introducing a Lagrange multiplier $P$, interpreted as a pressure. Consequently, the total energy functional $\mathcal{E}[\mathbf{r},\boldsymbol{\rho},P]$ of the system  is given by:

\begin{eqnarray}
\mathcal{E}[\mathbf{r},\boldsymbol{\rho},P,] & = & \frac{\mu}{2} \int_{\mathcal{B}_0} (\mbox{Tr}(\mathbf{F}^T\mathbf{F})-3) \mbox{d}V  + \gamma_s \int_{\partial \mathcal{B}} \mbox{d}a \nonumber
  \\
&&- \int_{\mathcal{B}_0} P ( \mbox{det}\, \mathbf{F} -1) \mbox{d}V - \int_{\partial \mathcal{B}'} \mathbf{f} \cdot  \mathbf{u} \mbox{d}a'
\end{eqnarray}

\noindent where $\boldsymbol{\rho}=\left\{\rho,d \right\}$ is the position of the contact line, $\mbox{d}V$ is an infinitesimal volume in the reference configuration while $\mbox{d}a$ (resp. $\mbox{d}a'$) is an infinitesimal element of area in the current (resp. prestretched) configuration $\mathcal{B}$ (resp. $\mathcal{B'}$). The vector $\mathbf{f}$ describes the force distribution applied at the free surface of the elastic layer by the drop. For a liquid hemispherical drop with radius $\rho$, surface energy $\gamma_{\ell}$ and macroscopic contact angle $\alpha$, the surface force distribution has two contributions: a localized traction $\mathbf{f}^{T} = \gamma_{\ell}  \delta(r'-\rho)(\sin{\alpha}\bold{e}_z -  \cos{\alpha}\bold{e}_r)$  at the triple line and a distributed compression $\mathbf{f}^{C} =-\gamma_{\ell} \sin{\alpha} /\rho \Pi(\rho - r')\bold{e}_z$ below the drop due to the Laplace pressure inside the drop, with $\Pi$ being the Heaviside function. 

\bigskip

\noindent \textbf{Variational procedure.} At equilibrium, the principle of stationary potential energy states that the variation of energy $\delta \mathcal{E}[\mathbf{r},\boldsymbol{\rho},P]$ with respect to small variations in the independent fields must be zero. This principle yields the equilibrium equations which must be completed by appropriate boundary conditions. Motivated by the experimental setups mentioned previously \cite{xu17,schulman18}, we assume that the lower surface of the elastic layer is bonded to an infinitely rigid surface, and thus $u_r(r',0)=u_z(r',0)=0$. From the definition of the energy functional above, the first Piola-Kirchoff tensor $\mathbf{P}$ is found to be $\mathbf{P} = \mu \mathbf{F} - P\mathbf{F}^{-1}$. Recalling that $\bf{F'} = \partial \bf{r'} / \partial \bf{R}$, it can be shown that the equilibrium equation can be written as $\mbox{div}(\mathbf{P} \bf{F'} )= 0$ where the $\mbox{div}$ operator is evaluated in the prestretched configuration $\mathcal{B'}$. Everywhere at the free boundary $z'=h$, except at the triple line, use of Nanson's formula gives $\mathbf{P} \mathbf{F'} \cdot \mathbf{n}' = \mathbf{f}^{C}  + \gamma_s \mathbf{n} \cdot(\mathbf{\nabla}\mathbf{n})$ where $\mathbf{n}'=(0,1)$ is the outward unit vector normal to the free surface in $\mathcal{B'}$ and $\mathbf{n}$ is the outward unit vector normal the free surface in $\mathcal{B}$. We now turn to the balance of forces at the triple line which follows from the variation of the energy with respect to $\boldsymbol{\rho}$. In the radial direction $\bold{e}_r$, this balance is purely configurational because the contact line position is a massless material point that is free to move (in absence of hysteresis) and thus \cite{gelfand00,podio01,gupta08}:

\begin{equation}
-\gamma_{\ell} \cos{\alpha} = \gamma_s \left\{ \cos{\theta^-} -\cos{\theta^+} \right\} + \bold{e}_r \cdot \mathbf{f}^{E}  \label{eq:BCTL2}
\end{equation}

\noindent where $\theta^- = \vert \partial u_z/\partial r (\rho^-,0) \vert$ and  $\theta^+ = \vert \partial u_z/\partial r (\rho^+,0) \vert$ are the (positive) angles of the solid on each side of the triple line. Because of this jump in the first derivative of the displacement field ($\partial u_z/\partial r (\rho^-,0) \neq \partial u_z/\partial r (\rho^+,0)$), which induces a logarithmic divergence of the stress, the tip of the elastocapillary ridge is a singular line or, in the language of Eshelbian mechanics, a \textit{defect}. More specifically, the ridge is a \textit{disclination} \cite{degennes93}. By contrast with disclinations in crystals however, the strength of the disclination, which is given by $1/2-\theta/2\pi$, can take any value between -1/2 and 1/2 as it is not related to an underlying lattice structure. The last term on (\ref{eq:BCTL2}) is precisely the well-known Eshelby force $\mathbf{f}^{E}$ acting on an elastic singularity  \cite{eshelby51,eshelby75}:

\begin{equation}
\mathbf{f}^{E} = \int_{\Gamma} (\mathcal{W}_e \mathbf{I}  - \mathbf{F}^T \mathbf{P})\boldsymbol{\nu} \mbox{d}\ell
\label{eq:eshelby}
\end{equation}

\noindent where $\Gamma$ is an arbitrary surface enclosing the defect and $\boldsymbol{\nu}$ is the outward unit normal vector to the surface $\Gamma$.  It has the dimensions of a force per unit length and is also called the J-integral in the context of fracture mechanics \cite{rice85}. The formula (\ref{eq:BCTL2}) is a new generalized law for contact lines in which the last term of the r.h.s is a line tension (per unit length) of elastic origin.  Note that this configurational force balance at the triple line is akin to the Erdman-Weierstrass condition that a broken extremal must satisfy at each corner point \cite{abonyi78}. In the vertical direction $\bold{e}_z$, the force balance also involves the elastic stresses, in all generality, because the vertical position of the contact line is constrained to lie on the free surface, i.e $u_z(\rho,0) = d$:

\begin{equation}
\gamma_{\ell} \sin{\alpha} = \gamma_s \left\{ \sin{\theta^-} +\sin{\theta^+} \right\} + \bold{e}_z \cdot \mathbf{f}^{E} + \lim_{\epsilon \rightarrow \infty} \int_{-\epsilon}^{\epsilon} \bold{e}_z \cdot  \mathbf{P}  \mathbf{F'} \cdot \mathbf{n}' \mbox{d}r \label{eq:BCTL1} 
\end{equation}

Some interesting limiting cases can readily be obtained from the force balances (\ref{eq:BCTL2}) and (\ref{eq:BCTL1}). In the case of a fluid at rest, the Eshelby force $\mathbf{f}^{E}$ vanishes \cite{gurtin00}. This follows from the fact that fluids are described using the current (deformed) configuration as the reference configuration (thus $ \mathbf{F} = \mathbf{I}$) and the first Piola-Kirchoff stress tensor reduces to the Cauchy stress which, at rest, is just a pressure $\mathbf{P} = -p \mathbf{I}$. The contour integral (\ref{eq:eshelby}) is thus $0$ and one recovers the Neumann construction that rules the equilibrium at triple lines between fluids. The situation is different for fluids in motion because the shear stress will induces configurational forces at the triple line \cite{gurtin00}. When the substrate is infinitely rigid on the other hand, the angles $\theta^-$ and  $\theta^+$ vanish. Equation (\ref{eq:BCTL2}) thus reduces to a generalized Young equation with line tension \cite{rusanov77} while (\ref{eq:BCTL1}) indicates that the vertical surface traction is solely balanced by the elasticity of the substrate for hard materials \cite{dervaux15}. 

\begin{figure*}[t!]
\begin{center}
\includegraphics[width=\textwidth]{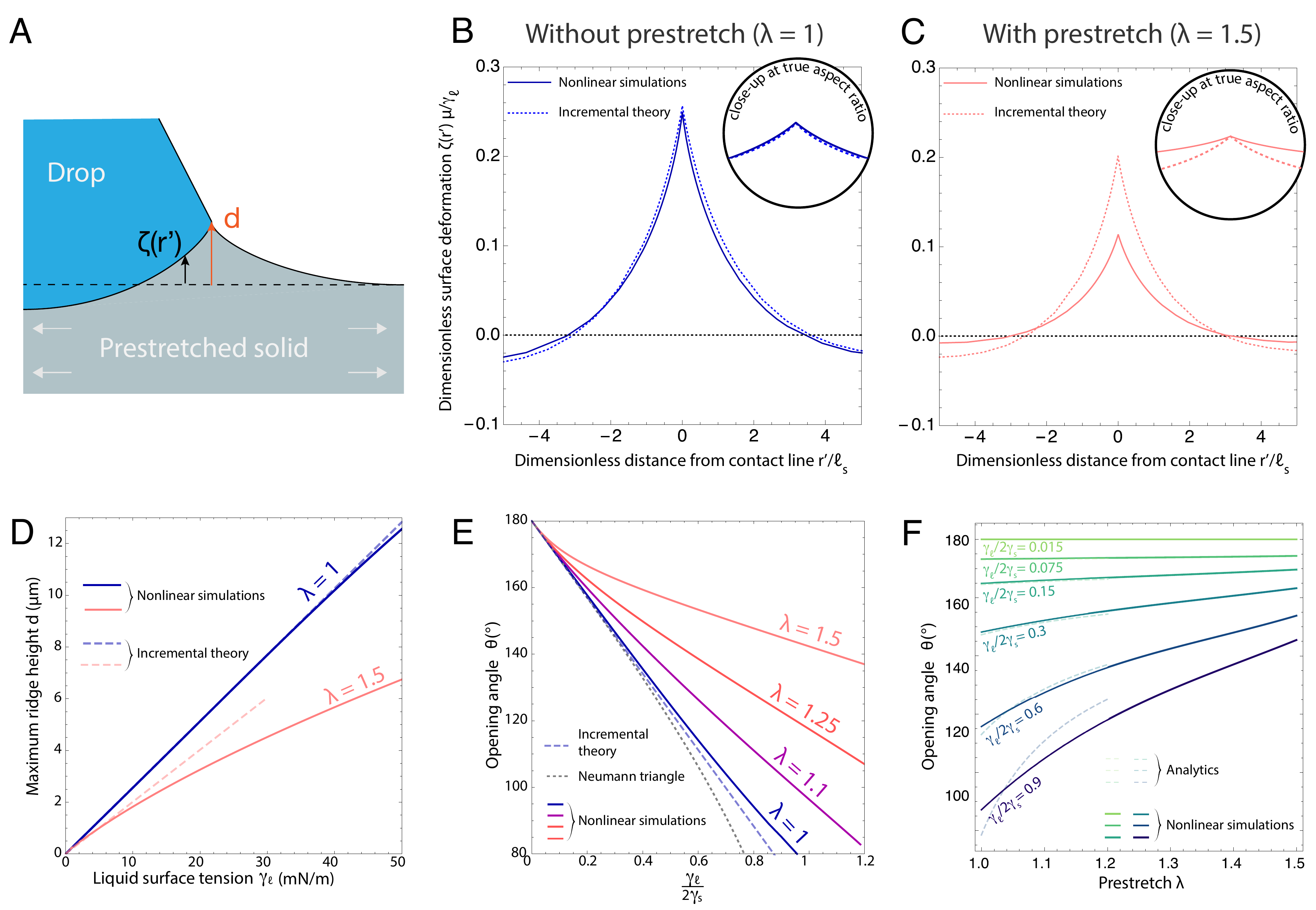}
\caption{{\footnotesize Results from the numerical simulations and comparison with the linear theory. A: schematic representation of the structure of the ridge with height $d$ and profile $\zeta(r')$. B and C: Structure of the elastocapillary ridge below the contact line without (resp. with) prestretch in panel B (resp. C). The solid lines are the results from the numerical simulation while the dotted lines show the linear theory. For all simulations presented in this paper, the initial thickness of the elastic layer is $80\mu$m to remain close to a typical experimental setup. The macroscopic contact angle $\alpha$ is taken to be $\pi/2$ and the radius of the drop is large (drop volume $5\mu$L and $\rho \sim 1.33$mm) in order to minimize the influence of the finite size of the drop and focus mostly on the possible effects of the prestretch and the nonlinearities. The height is normalized by $\gamma_{\ell}/\mu$ and the coordinate $r'$ is normalized by $\ell_s = \gamma_s/2\mu$. The insets in B and C show the detailed structure of the ridges (over a total width of $\ell_s$) at a true aspect ratio (i.e same normalization by $\ell_s$ for height and width). The linear model and the numerical simulations have been shifted vertically to allow for a better comparison between the two. D: Maximum ridge height, in microns, as a function of the liquid surface tension $\gamma_{\ell}$ for a solid surface tension $\gamma_{s} = 30$mN/m. E: opening angle as a function of the ratio $\gamma_{\ell}/2\gamma_s$ for different values of the prestretch. The blue dashed line is the prediction from the linear theory $\theta = \pi - \gamma_{\ell}/\gamma_s$ while the black dotted line is the results from the Neuman construction $\theta = \pi - 2\mbox{arcsin}(\gamma_{\ell}/2\gamma_s)$. F: opening angle as a function of the prestretch $\lambda$ for various values of the ratio $\gamma_{\ell}/2\gamma_s$. The light dashed lines are the results from the nonlinear approximation (\ref{eq:opening angle}).}}
\label{fig:mainfig}
\end{center}
\end{figure*}

\begin{figure*}
  \begin{minipage}[t]{0.74\textwidth}
 \vspace{0pt}
    \includegraphics[width=\textwidth]{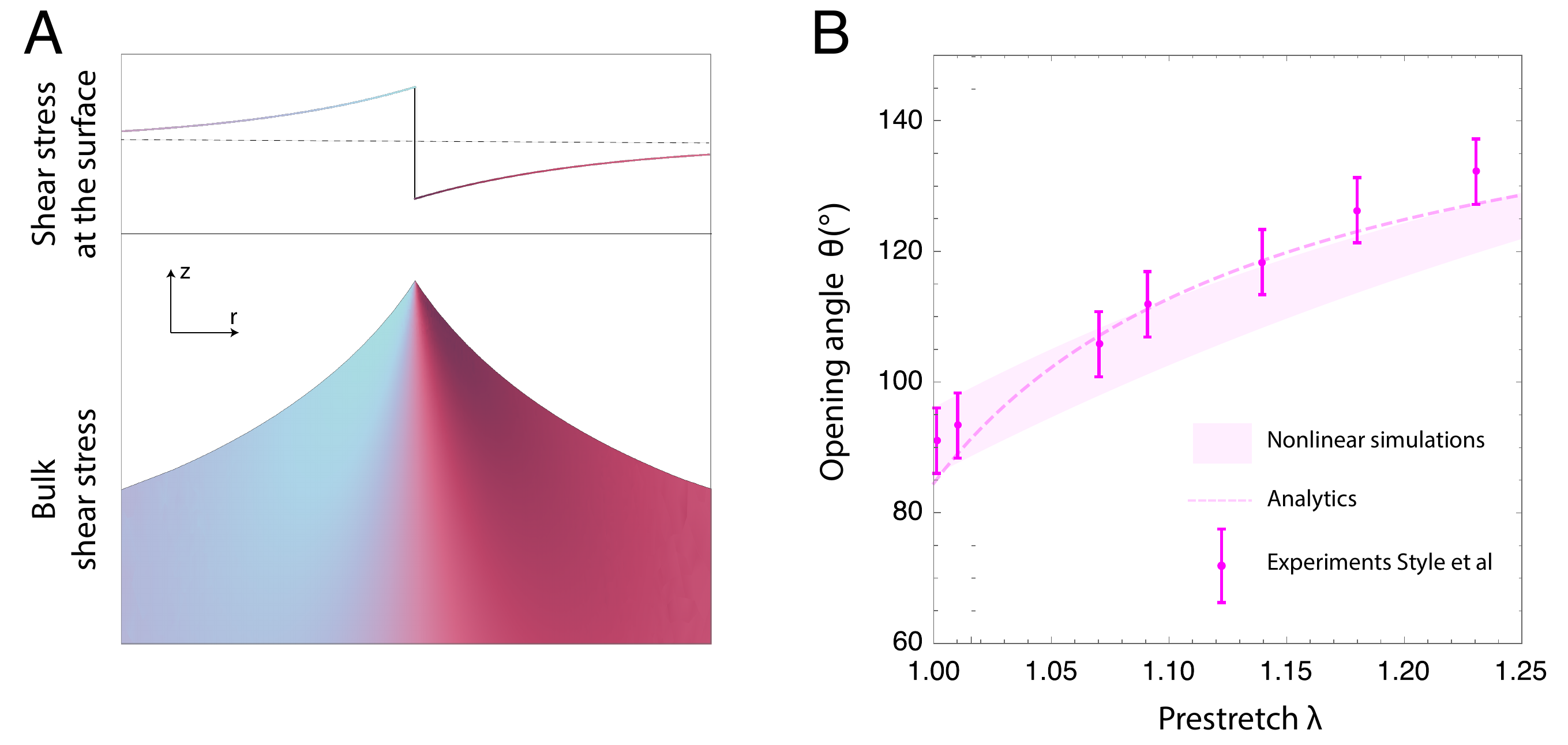}
  \end{minipage}\hfill
  \begin{minipage}[t]{0.25\textwidth}
  \vspace{0pt}
\caption{{\footnotesize A: Color-coded shear stress distribution in the ridge below the triple line obtained from the numerical simulations ($\lambda=1.2$, $\gamma_{\ell}/\gamma_s = 0.8$ and other parameters as specified in the caption of Fig.\ref{fig:mainfig}). The dimensions of the region shown here is $2\ell_s \times 2\ell_s$. The upper part of the panel shows the jump of the shear stress at the free surface across the contact line. B: comparison between the experimental data of \cite{xu17} for the opening angle of the ridge as a function of the prestretch and the numerical simulations as well as formula (\ref{eq:opening angle}).}}
\label{fig:fig3}
  \end{minipage}
\end{figure*}

These nonlinear equations are then solved numerically using a method we developed previously \cite{depascalis18} and the results are presented in Fig.\ref{fig:mainfig} and \ref{fig:fig3}. But before discussing these numerical data, let us first derive some analytical results that can be obtain in the limit where the displacement field $\bf{u}$ is small, i.e its amplitude is of order of a small parameter $\epsilon$, in which case it is referred to as an \textit{incremental deformation field}.

\bigskip

\noindent {\bf Incremental solution.} We now linearize the equilibrium condition and boundary conditions \textit{around the finitely prestretched configuration} in order to obtain the incremental solution \cite{ogden84}. To this end, it is convenient to separate the pressure into the form $P=P^{\star} + \epsilon p$ where $P^{\star}$ is a constant pressure in $\mathcal{B}'$ and $p$ is the incremental pressure from $\mathcal{B}'$ to $\mathcal{B}$. By applying the boundary condition  $\mathbf{P} \bf{F'} \vec{n}' = 0$ which holds in $\mathcal{B}'$ before the deposition of the drop, it is easily found that $P^{\star} = \mu/\lambda^4$. Following this decomposition, the equilibrium equations and boundary conditions can be linearized with respect to $\epsilon$ and are solved easily by shifting into Fourier space, leading to the following solution for the vertical displacement of the free surface $\zeta(r') = u_z(r',h)$:

\begin{equation}
\zeta(r') = \gamma_{\ell} \sin{\alpha} \int_0^{\infty} \mbox{d}s \frac{ J_0(sr')  \left(\rho J_0(s\rho) -\frac{2 J_1(s\rho)}{s}  \right)}{2\mu g(h s,\lambda)+ s\gamma_s} 
\label{eq:solution}
\end{equation}



\noindent where the function $g(hs,\lambda)$ is given in appendix. Note that when there is no prestretch, i.e when $\lambda =1$, the solution (\ref{eq:solution}) reduces to the well know solution of the elastowetting problem where the surfaces force distribution is damped both by the elasticity and the surface energy of the substrate. In absence of surface tension, we recover a classical result of incremental elasticity \cite{he08}. Interestingly, it should be noted that the introduction of a finite prestretch only affect the elastic term $2\mu g(h s,\lambda)$ but not the term associated with the surface energy of the solid $s\gamma_s$. This result could in fact be anticipated because the cost of creating a unit of area is independent of the underlying deformation in the absence of any Shuttleworth effect, and thus the surface energy term is independent of $\lambda$, as seen in (\ref{eq:solution}). Furthermore, it is known that the incremental response of a prestretched elastic half-space is that of a transversely isotropic linear half-space. Indeed, for very thick sample, i.e in the limit $h \rightarrow \infty $, we find that $g_{\infty}(\lambda) = \lim_{h \rightarrow \infty} g(h s,\lambda) = (\lambda^9+\lambda^6+3\lambda^3-1)/(2(\lambda^7+\lambda^4))$, implying that the incremental response of a prestretched Neo-Hookean elastic half-space with elastic modulus $\mu$ is identical to that of a linear elastic half-space without prestretch but with an effective shear modulus $\mu g_{\infty}(\lambda)$. This increase of the apparent rigidity is a purely nonlinear effect. As a consequence, the incremental deformation theory predicts that: i) the overall profile of the ridge, and in particular its height $d = \zeta(R)$, depend on the prestretch $\lambda$; ii) regarding the opening angle of the ridge however, we recover the classical result $\theta = \pi - 2\gamma_s \zeta'(0^{-}) = \pi - \gamma_{\ell}/\gamma_s$ for all thicknesses $H$ and prestretch $\lambda$. Therefore the incremental theory predicts that the opening angle is constant, for any prestretch $\lambda$, in the limit of small deformations ($\gamma_{\ell}/2\gamma_s \ll1$). 

\bigskip

\noindent {\bf Results.} Two deformation profiles given by the analytical solution (\ref{eq:solution}), as well as their height $d$,  are plotted in Fig.2 B-D, together with the results from the numerical simulation in absence of prestretch ($\lambda = 1$ in Fig.2-B) and in the case of an initial prestretch ($\lambda = 1.5$ in Fig.2-C) for $\gamma_{\ell}/2\gamma_s = 0.8.$  As shown in Fig.2-B the incremental theory provides a nice approximation to the numerical simulations of the nonlinear problem for the overall structure of the ridge, at both large ($r\gtrsim \ell_s = \gamma_s/(2\mu)$) and small ($r\lesssim \ell_s = \gamma_s/(2\mu)$) scale in absence of any prestretch ($\lambda = 1$). On the other hand, when the elastic layer is initially prestretched ($\lambda = 1.5$), the agreement between the incremental theory and the numerical simulations is poor as seen in Fig.2-C as the height of the ridge is smaller than expected from the incremental theory and has a broader opening angle. Focusing on the height of the ridge only (Fig.2-D) indicates that, in presence of prestretch, the numerical simulations coincide with the incremental solution only at small values of the ratio $\gamma_{\ell}/2\gamma_s$ but start deviating from the incremental theory very quickly. In absence of prestretch, the agreement is much better over the whole range of $\gamma_{\ell}/2\gamma_s$ investigated here. We now focus on the opening angle $\theta$ of the ridge which is the central observable allowing to conclude on the existence of the Shuttleworth effect. As seen in Fig.2-E, the opening angle of the ridge decreases with increasing value of the ratio $\gamma_{\ell}/2\gamma_s$, as expected from the linear theory, but surprisingly strongly increases with the prestretch, in opposition with the prediction from the linear theory. Furthermore, note that even in the case $\lambda=1$, the opening angle is always larger than predicted by the linear theory, with the difference increasing with the ratio $\gamma_{\ell}/2\gamma_s$. This difference is of course even more pronounced with the Neumann construction which fails to predicts the opening angle of the ridge. Fig.2-F shows that the opening angle $\theta$ increases monotonously as a function of the prestretch $\lambda$ for various values of the ratio $\gamma_{\ell}/2\gamma_s$. 


\bigskip

\noindent \textbf{Discussion.} In the preamble of this paper, two questions were raised pertaining to the validity of the Neumann construction at finite deformation in the presence or absence of a prestretch $\lambda$. The results presented in this study allow us to provide answers to these questions.

In absence of any pre-stretch ($\lambda=1$), and for values of the ratio $\gamma_{\ell}/2\gamma_s$ up to $\sim 0.9$, the linear theory $\theta = \pi - \gamma_{\ell}/\gamma_s$ is a reasonable approximation to the numerical solution of the fully nonlinear elastowetting problem, with less than $5\%$ relative difference between the two models for both the opening angle $\theta$ of the ridge as well as the height of the ridge. The agreement is not as good however with the Neumann construction $\theta = \pi - 2\mbox{arcsin}(\gamma_{\ell}/2\gamma_s)$. More precisely, for a value of the ratio $\gamma_{\ell}/2\gamma_s \sim 0.9$, typical of experiment, the predictions for the opening angle between the Neumann construction and the nonlinear simulations differ by roughly $30^{\circ}$. As this difference is much larger than the precision of typical experimental measurements, we may answer no to the first question "\textit{does the Neumann construction hold at experimentally relevant finite value of $\gamma_{\ell}/2\gamma_s$ ?}  The situation is even worse for a drop deposited on a prestretched elastic layer. In this case, the numerical simulation starts deviating from the incremental theory, even at very small deformations  $\gamma_{\ell}/2\gamma_s \ll 1$. In particular, and this is the main result of this work, the simulations predict that the opening angle $\theta$ of the ridge is an increasing function of the prestretch $\lambda$. This dependence is a pure nonlinear effect as both the incremental theory as well as the Neumann construction predict that $\theta$ does not depend on $\lambda$. To the second question "\textit{does the Neumann construction hold when the elastic layer is initially prestretched before the deposition of the drop, at experimentally relevant finite values of  $\gamma_{\ell}/2\gamma_s$ ?}", the answer is therefore also no. 

While a full analytical solution of the nonlinear problem is out of reach at present, some analytical progress can nonetheless be made to understand more quantitatively the curve $\theta(\lambda)$ shown above. Because the stress field around the elastocapillary ridge is equivalent to that around a wedge disclination, it can be shown within the framework of linear elasticity  \cite{mura70,vitek74,eshelby82} that the Eshelby force (\ref{eq:eshelby}) for a disclination line in an external stress field  can be calculated  as: $\mathbf{f}^{E} \approx - 2 \bold{e}_{\theta} \times (2\pi S \bold{M} \cdot \bold{e}_{\theta})$. This force on a disclination is a direct analog of the Peach-Koehler force \cite{peach50} acting on a dislocation. The factor of $2$ is due to the presence of the free surface that acts as a mirror disclination of opposite strength $-S$.   Here $ M_{jm} = T_{ji} \epsilon_{imn}u_n(R)$ is the torque on the defect, $T_{ji}$ is the Cauchy stress, $\epsilon_{imn}$ is the Levi-Civita tensor and Einstein summation convention applies. At leading order, the vertical component of the Eshelby force on acting on the ridge is thus:

\begin{equation}
f^{E}_z  \approx - 4 \pi S T_{rr}^{(0)} \zeta(R) = -2 \mu (\pi - \theta)(\lambda^2-\frac{1}{\lambda^4})\zeta(R)
\label{eq:eshelbyfinal}
\end{equation}

The analogy with the Peach-Koehler force is even more obvious in formula (\ref{eq:eshelbyfinal}). Indeed, the vertical component of the Peach-Koehler force on a surface dislocation simply reads $2 [ T_{rz} u_z ] $ where the bracket operator $[ f ]$ denotes the jump of f across the defect. In the case of a dislocation, the stress field itself is continuous while the jump of the displacement $u_z$ is non-zero (and defined as the Burger vector). In our case the displacement $u_z$ is continuous while the shear stress is discontinuous  at the triple line (as seen in Fig.\ref{fig:fig3}-A) as it follows from the boundary condition at the free surface that $T_{rz} = -T_{rr}^{(0)}\partial \zeta/\partial r'$. Using this expression in the Peach-Koehler formula, one immediately recovers expression (\ref{eq:eshelbyfinal}). A rather interesting feature of the force (\ref{eq:eshelbyfinal}) is that is it essentially independent of the elastic modulus because the height of the ridge is inversely proportional to the substrate shear modulus $\zeta(R) = a(R,H) \gamma_{\ell} \sin{\alpha}/(\mu g_{\infty}(\lambda))$. Here $a(R,H)$ is simply a geometric parameter that is weakly dependent on the thickness $H$ and the droplet size $R$, provided that both are larger than the elastocapillary length $\ell_s$, and whose value is roughly $\sim 0.25$. Thus we have the approximation $f^{E}_z \approx -  \gamma_{\ell}\sin{\alpha}/(2 g_{\infty}(\lambda)) (\lambda^2-1/\lambda^4)(\pi -\theta)$. It is then easily seen that the Eshelby force $f^{E}_z $ as the same effect on the triple line as a surface energy of magnitude $ \gamma_{\ell}\sin{\alpha}/(2 g_{\infty}(\lambda)) (\lambda^2-1/\lambda^4)$. Consequently, we may define an "apparent surface tension" $\Upsilon$ given by:

\begin{equation}
\Upsilon  \approx \gamma_s \left\{1+ \frac{\gamma_{\ell} \sin{\alpha}}{\gamma_s} \frac{\lambda^9+\lambda^6-\lambda^3-1}{\lambda^9+\lambda^6+3\lambda^3-1} \right\}
\label{eq:apparentsurfacetension}
\end{equation}

\noindent which at small $\lambda$ has the following simple approximation: $\Upsilon  \approx \gamma_s \left\{1+ 3\frac{\gamma_{\ell} \sin{\alpha}}{\gamma_s}(\lambda-1) \right\}$. Furthermore, equation (\ref{eq:apparentsurfacetension}) leads to the following approximation for the opening angle of the ridge:

\begin{equation}
\theta  \approx \pi - \frac{\gamma_{\ell}}{\Upsilon}
\label{eq:opening angle}
\end{equation}

Note that the expressions above are based on a very crude approximation of the Eshelby force because we have neglected here the self-force of the disclination on itself as well as the force induced by the Laplace pressure on the defect.  Although these contributions are higher-order contributions than the leading term presented in equation (\ref{eq:eshelbyfinal}), they can become significant when $\gamma_{\ell}/2\gamma_s$ is of order unity. Nonetheless, as seen in Fig.(\ref{fig:mainfig})-F, the  formula (\ref{eq:opening angle}) already provides a reasonable approximation for the opening angle of the ridge. Interestingly, it should be noted that the existence of an elastic restoring force, which magnitude is proportional to the height of the ridge and to the shear modulus of the substrate, as in $\ref{eq:eshelbyfinal}$, was very recently reported in corse-grained molecular dynamics simulations \cite{liang18}.

We now compare those theoretical predictions to available experimental data. On an unstretched PDMS substrate the opening angle of the ridge was measured by \cite{xu17} as $91.2^{\circ}$ for a glycerol droplet ($\gamma_{\ell}=41\pm1$mN/m). According to the Neumann construction, this yield a solid surface energy of $29$mN/m, that is $\sim 40\%$ larger than the surface energy of liquid PDMS ($21\pm1$mN/m). According to the nonlinear simulations on the other hand, such an opening angle yields a surface energy of $24$ mN/m, much closer to the surface energy of liquid PDMS. Turning now to the dependence $\theta(\lambda)$, it can be seen in Fig.\ref{fig:fig3}-B that the numerical simulations, \textit{in absence of any Shuttleworth effect}, reproduce the experimental data of \cite{xu17}. From this nice agreement, we may conclude that the experimental observations of \cite{xu17} are essentially a consequence of the Peach-Koehler force acting on the elastocapillary ridge and that, within the experimental error bars, there is no need to assume a Shuttleworth effect in soft elastomers, in complete agreement with the experimental data of \cite{schulman18}.  In addition, we also show in Fig. \ref{fig:mainfig}-F and Fig. \ref{fig:fig3}-B the prediction given by the analytical formula (\ref{eq:opening angle}) as a function of the prestretch $\lambda$, which shows a very nice agreement with the numerical and experimental data. It is important to note however, that the apparent surface tension defined in equation (\ref{eq:apparentsurfacetension}) only appears in the force balance at the tip of the ridge as a consequence of the corner singularity and cannot be used as a pseudo-Shuttleworth effect that would apply everywhere at the surface of the elastic domain.  Finally, let us comment on a few remarkable behaviors of equation (\ref{eq:apparentsurfacetension}). First, this expression predicts that the effective surface tension decreases under compression and vanishes at $\lambda \approx 0.82$. Second, between this critical value and the critical stretch of the Biot instability $\lambda \approx 0.666$ \cite{biot63}, the effective surface tension is \textit{negative} because the Peach-Koehler force exceeds the restoring effect of the solid surface energy. This region of the phase space would therefore be an interesting regime to explore in experiments. 

To conclude, we have unraveled in this study a new general balance of forces (\ref{eq:BCTL2})-(\ref{eq:BCTL1}) ruling the behavior of contact lines on soft materials. This conceptual breakthrough has shed new light on various highly debated issues in elastowetting and shown the failure of the Neumann construction for liquid drops on solid surfaces as well as the absence of the Shuttleworth effect in elastomers.  Our approach will likely help understand and control the complex interactions between drops on soft surfaces. Indeed, defects such as disclinations repel each other when they have strengths $S$ of the same sign and our approach might help predict in a quantitative, original and simple way the interactions between droplets. Furthermore, while we have focused in this study on the static of elastowetting, some interesting phenomena might also arise in dynamical elastowetting \cite{karpitschka15,zhao18a,zhao18b} because Peach-Koehler forces are additional source of dissipations beside classical visco-elastic stresses. Moving beyond the elastowetting problem studied in this work, we also anticipate that the theoretical framework developed here will provide a valuable tool to understand complex physical phenomena related to the formation of singular structures in elasticity, such as the long-standing issue of cusp formation arising in the Biot instability \cite{dervaux11,dervaux11b,hohlfeld11,karpitschka17}.


\noindent \textbf{Acknowledgements} 

ANR (Agence Nationale de la Recherche) and CGI (Commisariat \`{a} l'Investissement d'Avenir) are gratefully acknowledged for their financial support through the GELWET project (ANR-17- CE30-0016), the Labex SEAM (Science and Engineering for Advanced Materials and devices - ANR 11 LABX 086, ANR 11 IDEX 05 02) and through the funding of the POLYWET and MMEMI project.


\noindent \textbf{Appendix}

\begin{widetext}
\begin{equation}
g(h s,\lambda) = \frac{-\left(\lambda ^{12}+6 \lambda ^6+1\right) \sinh (h s) \sinh \left(h \lambda ^3 s\right)+\left(\lambda ^{12}+2 \lambda ^6+5\right) \lambda ^3 \cosh (h s)
   \cosh \left(h \lambda ^3 s\right)-4 \left(\lambda ^9+\lambda ^3\right)}{2 \lambda ^4 \left(\lambda ^6-1\right) \left(\lambda ^3 \sinh (h s) \cosh \left(h \lambda
   ^3 s\right)-\cosh (h s) \sinh \left(h \lambda ^3 s\right)\right)}
\end{equation}
\end{widetext}

\end{document}